\documentclass[aip,amsmath,amssymb,preprint]{revtex4-1}

\usepackage{graphicx}

\begin{document}

\title{\textit{Ab-initio} Modeling of CBRAM Cells: from
  Ballistic Transport Properties to Electro-Thermal Effects}

\author{F.~Ducry$^{*}$, A.~Emboras$^{\dagger}$, S.~Andermatt$^{*}$,
M.~H.~Bani-Hashemian$^{*}$, B.~Cheng$^{\dagger}$,
J.~Leuthold$^{\dagger}$, and M.~Luisier$^{*}$}

\affiliation{$^*$Integrated Systems Laboratory (ETH Zurich) /
  $^{\dagger}$Institute of Electromagnetic Fields (ETH Zurich) 8092 Zurich, Switzerland}

\begin{abstract}
We present atomistic simulations of conductive bridging random access
memory (CBRAM) cells from first-principles combining
density-functional theory and the Non-equilibrium Green's Function
formalism. Realistic device structures with an atomic-scale filament
connecting two metallic contacts have been constructed. Their transport
properties have been studied in the ballistic limit and in the
presence of electron-phonon scattering, showing good agreement with
experimental data. It has been found that the relocation of few atoms
is sufficient to change the resistance of the CBRAM by 6 orders of
magnitude, that the electron trajectories strongly depend on the
filament morphology, and that self-heating does not affect the device
performance at currents below 1 $\mu$A. 
\end{abstract}

\maketitle 

% ================== INTRODUCTION =============================================
\section{Introduction}
The \textit{I-V} characteristics and non-volatile storage capability
of conductive bridging random access memories (CBRAM) strongly depend
on the atomic properties of the underlying nano-filaments that form
between two metallic plates through a dielectric layer
\cite{waser}. Due to the extremely narrow dimensions of these 
filaments, high current densities are expected, with potentially 
significant self-heating effects. To design better performing CBRAM
cells it is therefore critical to precisely understand the interplay
between electron transport and atom positions as well as their
influence on temperature.

Device simulations can address this issue and give insight into
the functionality of CBRAMs, provided that the following two
modeling challenges are resolved. First, tools that can handle
realistically extended structures rely on classical physics, they do
not capture the atomic granularity of filaments, and they require
several material parameters as inputs \cite{ielmini}. Secondly, quantum
mechanical solvers based on \textit{ab-initio} methods do not
suffer from these limitations, but they are restricted to systems made
of a couple hundred atoms, either with ballistic transport or simple
scattering approaches \cite{nakamura}.

Advanced physical models that combine density-functional theory (DFT) 
\cite{dft} and the Non-equilibrium Green's Function (NEGF) formalism 
\cite{negf} have been developed here to eliminate these bottlenecks
and demonstrate the first atomistic quantum transport calculations of
realistic CBRAMs, as depicted in Fig.~\ref{fig:1}. For that purpose,
structures with up to 4500 atoms have been carefully constructed and 
simulated in the ballistic limit and with coupled electron-phonon 
transport. These breakthroughs have been the key to obtain results in
good agreement with experiments, to identify possible electron
trajectories through the studied device, to shed light on the
nano-filament dissolution process, to assess the power dissipated in 
the cell, and to determine the lattice temperature of each atom.

\section{Approach}\label{sec:app}
\begin{figure}
\centering
\includegraphics[width=\linewidth]{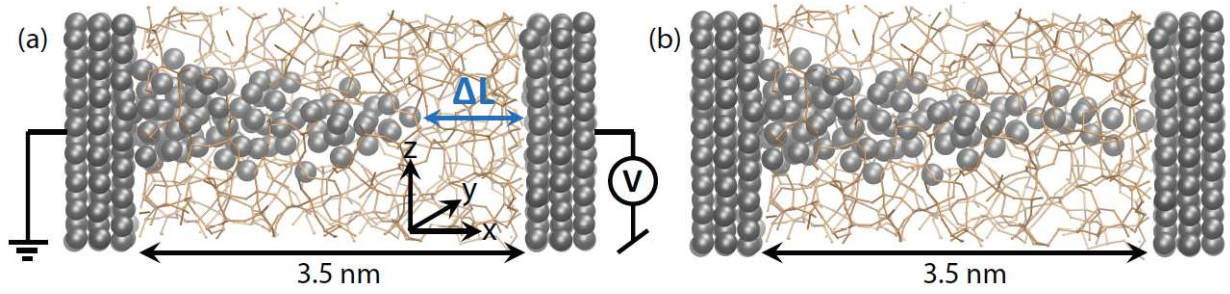}
\caption{Schematic view of the 3-D atomic structures considered in this
  work. Contrary to the experimental device, they are composed of
  two 4.25 nm long, $<$111$>$-oriented copper metallic plates
  (instead of platinum on the left) surrounding a 3.5 nm long SiO$_2$
  matrix through which a Cu nano-filament growths and dissolves upon
  application of an external voltage $V$. The cross section along the
  $y$ and $z$ axes (assumed periodic) measures 2.34$\times$2.38
  nm$^2$, respectively. Each structure is composed of roughly 4500
  atoms, where the large gray spheres represent Cu atoms, the small
  orange ones either Si or O. (a) Incomplete nano-filament 
  with a gap $\Delta L$=1 nm between its end and the active Cu
  plate. (b) Full, cone-shaped nano-filament made of 96 Cu
  atoms.} 
\label{fig:2}
\end{figure}

Since the goal of this paper is not to simulate the growth of
metallic nano-filaments, but to study the transport properties of
CBRAM cells, the first modeling step consists in creating practical
device structures made up of two copper contacts separated
by a slab of amorphous silicon dioxide (a-SiO$_{2}$). This layer has
been generated by a melt-and-quench approach using classical
force-field molecular dynamics routines from the ATK tool \cite{atk}
and a $\beta$-cristobalite SiO$_2$ crystal of 1080 atoms as starting
point. The final configuration has been relaxed with DFT using the
CP2K package \cite{cp2k} before attaching Cu contacts to the
a-SiO$_{2}$ and inserting a conical nano-filament by
replacing the silicon and oxygen atoms with Cu. The resulting
structure has been further relaxed and annealed for several ps
with DFT. Two examples composed of about 4500 atoms are reported in
Fig.~\ref{fig:2}.

\begin{figure}
\centering
\includegraphics[width=\linewidth]{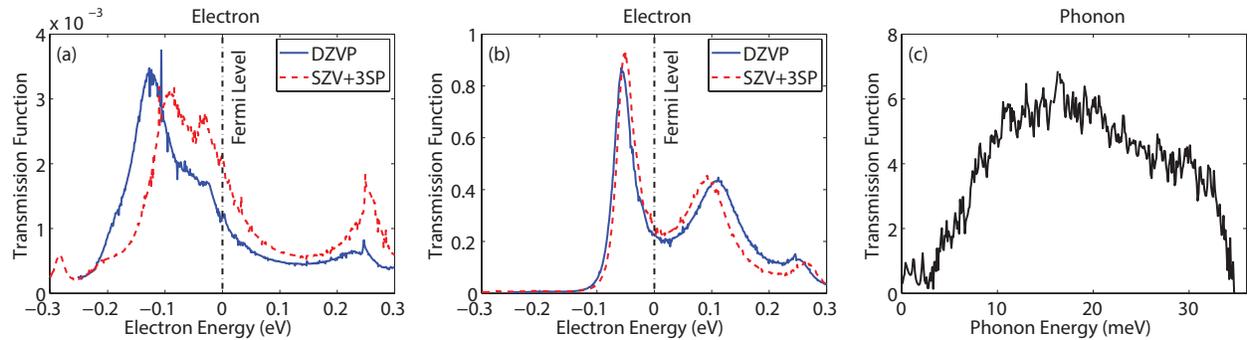}
\caption{(a) Energy-resolved electron transmission function $T(E)$
  through the incomplete nano-filament from Fig.~\ref{fig:2}(a) with
  $\Delta L$=1 nm and 9 atoms missing (87 instead of 96 Cu atoms). The
  reported data were computed with a Hamiltonian matrix created by
  CP2K and expressed in either a double-zeta valence polarized (DZVP,
  solid blue line) basis set or a combination of 3SP for the Si and O
  atoms and single-zeta valence (SZV) for the Cu ones (dashed red
  line). Note that the Fermi level energy was shifted to $E$=0. (b)
  Same as (a), but for the complete nano-filament from 
  Fig.~\ref{fig:2}(b). (c) Phonon transmission function through the
  same structure as in (b) with the dynamical matrix obtained from a
  reactive force field (reaxFF) model with the parameters from
  Ref.~\cite{onofrio}.} 
\label{fig:3}
\end{figure}

As next step, the Hamiltonian $H$ and overlap $S$ matrices of the
produced CBRAMs have been prepared by CP2K using contracted
Gaussian-type  orbitals as basis set, PBE exchange-correlation
functional \cite{gga}, and GTH pseudopotentials \cite{gth}. These
matrices have then been loaded into a quantum mechanical device
simulator \cite{omen} to compute the ballistic transport
characteristics of the corresponding structures. While double-zeta
valence polarized (DZVP) basis sets are known to be more accurate than
single-zeta valence (SZV) or 3SP ones, they induce much larger
computational burden. In Fig.~\ref{fig:3}(a-b) it is however shown
that for the considered cells, the electron transmission function as
calculated with a SZV+3SP combination agrees well with the one from
DZVP, thus justifying the usage of the lighter basis set.

Finally, to go beyond ballistic transport and account for
electro-thermal effects, the dynamical matrix (DM) of the systems in
Fig.~\ref{fig:2} has also been calculated using the frozen-phonon
approach of ATK and the reactive force-field (reaxFF) parameters of
Ref.~\cite{onofrio}. With the DM, the phonon transmission function can
be evaluated, as indicated in Fig.~\ref{fig:3}(c). More importantly,
if the derivatives of $H$ and $S$ are also built with CP2K, all
necessary quantities are available to solve the electron and phonon
Green's Functions and to self-consistently couple them via scattering
self-energies \cite{rhyner}. This has been done here to gain access to
the lattice temperature distribution inside atomic-scale CBRAM cells.

\section{Results}\label{sec:res}
\begin{figure}
\centering
\includegraphics[width=\linewidth]{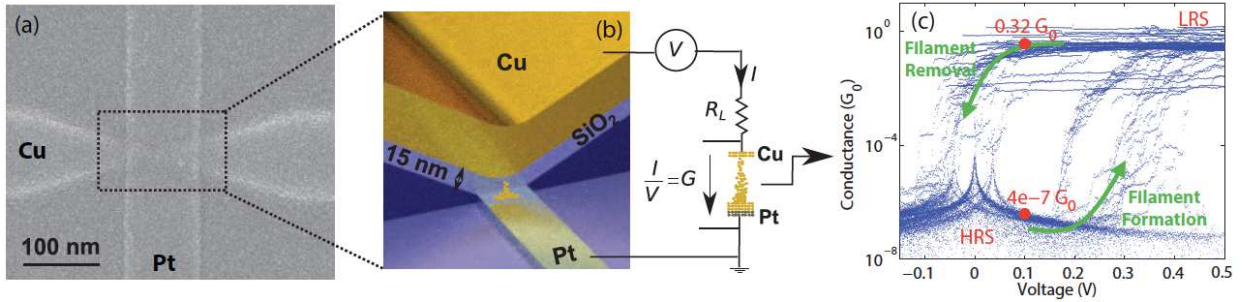}
\caption{(a) Scanning electron microscope view of a Cu-SiO$_2$-Pt
  conductive bridging random access memory (CBRAM) cell that
  was fabricated on a silicon-on-insulator wafer with a 220 nm thick
  silicon layer and a 3 $\mu$m-thick buried oxide. A 3D atomic point
  contact was created in the middle via local oxidation of silicon
  to precisely control the location of the forming metallic
  nano-filament. The Cu and Pt contacts were deposited using 
  a sequence of e-beam lithography, e-beam evaporation, and lift-off
  processes. (b) Illustration of the central active region of the CBRAM
  structure. The separation between the Cu and Pt layers can be made
  as thin as 15 nm and is filled with SiO$_2$. The device is put in
  series with a resistance to limit the current magnitude. A DC
  voltage is applied to the Cu contact, while the Pt one remains
  grounded. This bias triggers the growth of a Cu nano-filament
  through the SiO$_2$ matrix, starting from the Pt side. (c) Extracted
  conductance $G$ vs. voltage $V$ (as a function of the conductance
  quantum $G_0$) for the Cu-SiO$_2$-Pt CBRAM in (a) and
  (b). 30 set/reset hysteretic cycles of a single device are
  reported. The two red dots refer to the low and high resistance
  states of the structures shown in Fig.~\ref{fig:2}.}
\label{fig:1}
\end{figure}

\begin{figure}
\centering
\includegraphics[width=\linewidth]{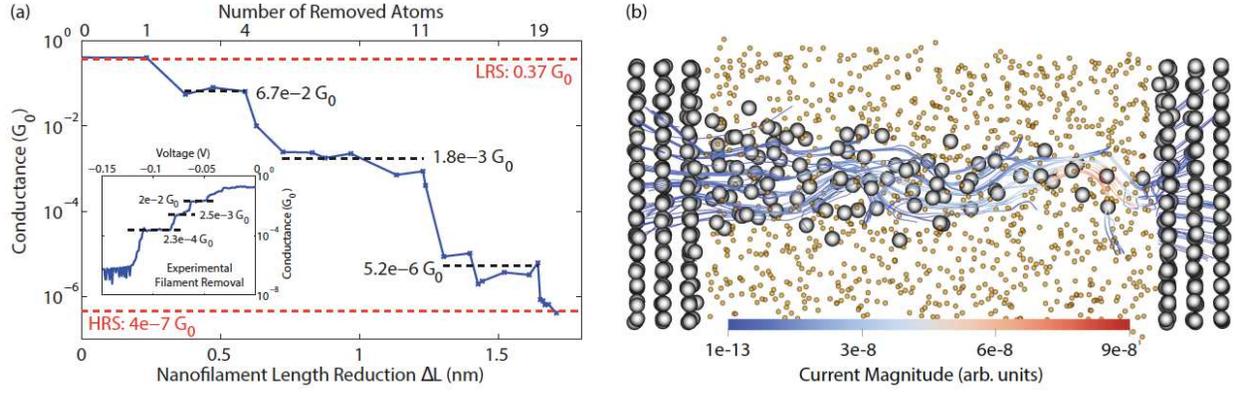}
\caption{(a) Ballistic conductance $G$ extracted from the CBRAM cell
  in Fig.~\ref{fig:2}(b) as a function of the number of Cu atoms
  removed from the filament extremity. By gradually
  expelling the far most right filament atom, a gap of length
  $\Delta$L, as sketched in Fig.~\ref{fig:2}(a), opens up and
  leads to a step-like reduction of $G$, as in experiments (see inset:
  zoom in of the area of filament removal in
  Fig.~\ref{fig:1}(c)). About 20 atoms must be removed  
  to go from the LRS ($G$=0.32 $G_0$) to the HRS ($G$=4e-7 $G_0$). (b) Ballistic
  current map through the same CBRAM as before under the application
  of an external voltage $V$=1 mV. The blue and red lines illustrate
  the electron trajectories, as computed from quantum transport, while
  the gray dots refer to Cu atoms, the orange ones, to Si and O.} 
\label{fig:4}
\end{figure}

First, the ballistic current flowing through the CBRAM in
Fig.~\ref{fig:2}(b) has been computed and is displayed in
Fig.~\ref{fig:4}(a) in the form of the conductance $G$. The obtained
low resistance state (LRS) value $G$=0.32 $G_0$ ($G_0$: conductance
quantum), falls exactly in the same range as the one measured from the 
experimental device in Fig.~\ref{fig:1}. This is a strong indication
that the technique used to generate the atomic systems gives rise to
realistic filament geometries. Since CBRAMs usually operate between a
low and high resistance state (HRS), we have tried to establish how
many Cu atoms should move away from the filament tip to enable a
LRS$\rightarrow$HRS transition. Starting from the structure in
Fig.~\ref{fig:2}(b), the far most right filament atoms have been
removed one-by-one and the conductance at each step simulated. The
results in Fig.~\ref{fig:4}(a) reveal that 20 atoms must be
relocated and a gap of 1.5 nm created to go from the LRS to the
HRS. Furthermore, the conductance of the decreasing nano-filament
exhibits several plateaus, in qualitative agreement with experiments
(see inset of Fig.~\ref{fig:4}(a)), suggesting that the removal of some
atoms is more critical than others. This is confirmed by the current
map reported in Fig.~\ref{fig:4}(b): it can be clearly seen that
electrons follow curly paths and avoid certain filament regions. As a
consequence, the presence or absence of atoms carrying little current
does not affect the overall device conductance.

\begin{figure}
\centering
\includegraphics[width=\linewidth]{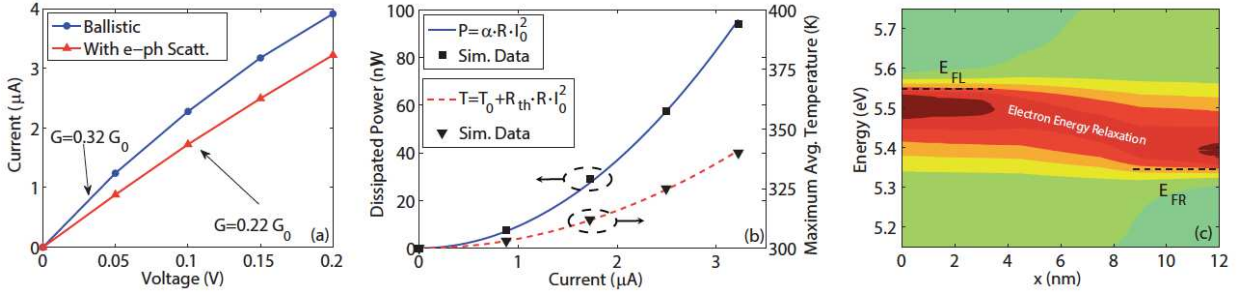}
\caption{(a) Current vs. voltage characteristics of the CBRAM cell in
  Fig.~\ref{fig:2}(b) in the LRS. The current in the ballistic limit
  (blue curve with circles) and as obtained with coupled
  electro-thermal transport (red curve with triangles) are reported,
  together with the corresponding conductance values. (b) Power
  dissipated inside the same structure as before 
  ($P_{diss}$, left $y$-axis) and average maximum lattice temperature
  ($T_{max}$, right $y$-axis) as a function of the electrical
  current $I_0$. The simulated power (black squares) follows a
  $P_{diss}$=$\alpha\cdot R\cdot I_0^2$ relationship (solid blue curve),
  where $\alpha$=0.16 is the amount of power dissipated inside the
  simulation domain (the rest is wasted in the contacts) and $R$=58
  k$\Omega$ is the LRS device resistance. The simulated temperature
  (black triangles, see Fig.~\ref{fig:6}(a) for the exact definition)
  exhibits a similar quadratic behavior
  $T_{max}$=$T_0$+$R_{th}\cdot R\cdot I_{0}^2$ (dashed red curve),
  with $T_0$=300 K and $R_{th}$=68 K/$\mu$W as the fitted thermal
  resistance \cite{ielmini}. (c) Spectral current (current as a
  function of position and energy) for the device in (a)
  and (b), in the presence of electron-phonon scattering ($V$=0.2
  V). Red indicates high current concentrations, green no 
  current. The left and right contact Fermi levels
  ($E_{FL}$ and $E_{FR}$) are indicated.}
\label{fig:5}
\end{figure}

The \textit{I-V} characteristics of the investigated CBRAM can be
found in Fig.~\ref{fig:5}(a) for the LRS up to $V$=0.2 V, both in the
ballistic limit and with coupled electron-phonon transport, assuming a
linear potential drop between the contacts. The ballistic
resistance/conductance is obviously not constant and changes as a
function of the applied voltage, contrary to the one obtained in the
presence of electron-phonon interactions. Turning on scattering
decreases the low voltage $G$ from 0.32 to 0.22 $G_0$. It can
therefore be estimated that the simulated device operates at about
70\% of its ballistic limit. Note that $G$=0.22 $G_0$ is still in the
experimental range (see Fig.~\ref{fig:1}).

By coupling electron and phonon transport, the power dissipated inside
the CBRAM can be accurately determined from the difference in thermal
energy current between the left and right contacts. An effective
lattice temperature can also be derived for each atom from the
non-equilibrium phonon population \cite{rhyner}. Collecting these data
requires that the electrical and energy currents are conserved, which
has been verified in the present case. The dissipated power $P_{diss}$
and maximum average temperature $T_{max,avg}$ are plotted in
Fig.~\ref{fig:5}(b) as a function of the current $I_0$: both
quantities behave as predicted by simpler physical models
\cite{ielmini}, i.e. they increase quadratically with the current. Up
to a current $I_0$=1 $\mu$A, $P_{diss}$ and $T_{max,avg}$ do
not exceed 10 nW and 305 K, respectively, so that self-heating is
almost negligible. Past this point, the situation rapidly
deteriorates as $I_0$ increases.

The simulated dissipated power obeys a $P_{diss}$=$\alpha\cdot R\cdot
I_{0}^2$ rule with the CBRAM LRS resistance $R$=58 k$\Omega$. The
pre-factor $\alpha$=0.16 indicates that only 16\% of the injected
electrical power is converted into heat inside the simulation domain,
the rest being wasted in the right contact and having a limited impact
on the nano-filament dynamics. The reason behind $\alpha<$1 can be
inferred from the spectral current distribution in
Fig.~\ref{fig:5}(c). At the applied voltage $V$=0.2 V, electrons 
should undergo an energy relaxation of 200 meV corresponding to the
Fermi level difference. However, the distance between both electrodes
is too short for electrons to emit enough phonons and 
dissipate the expected energy. 

\begin{figure}
\centering
\includegraphics[width=\linewidth]{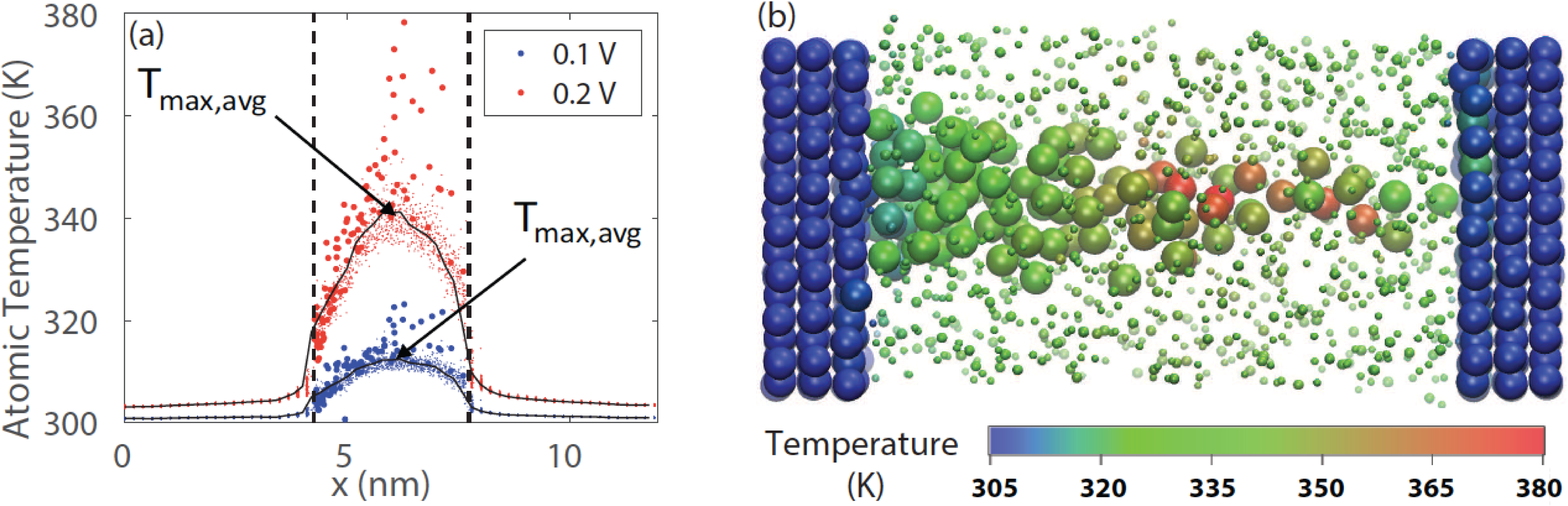}
\caption{(a) Atomically resolved lattice temperature as a function of 
  the $x$-coordinate of each atom for the device from
  Fig.~\ref{fig:2}(b) at $V$=0.1 (blue symbols) and 0.2 V (red
  symbols). The Cu atoms forming the central nano-filament are marked
  with larger circles. The thin black lines show the average lattice
  temperatures, as computed by averaging the temperature of all atoms
  present in a segment of length $dx$=0.2 nm along the $x$-axis. Their
  maximum, $T_{max,avg}$, are indicated and reported in
  Fig.~\ref{fig:5}(b). (b) Same as (a), but with the color of the
  atoms directly representing their lattice temperature at $V$=0.2 V.} 
\label{fig:6}
\end{figure}

The relationship between $T_{max,avg}$ and the electrical current
$I_0$ can be fitted with the following quadratic function
$T_{max,avg}$=$T_0$+$R_{th}\cdot R\cdot I_{0}^2$, where $T_0$=300 K is
the room temperature. A thermal resistance $R_{th}$=68 K/$\mu$W can be
extracted from the simulation data. In the fitting procedure, it is
crucial to consider the average temperature, and not individual values
because large variations may occur between the coldest and hottest
atoms situated in the same segment of length $dx$=0.2 nm along the $x$
direction, as can be observed in Fig.~\ref{fig:6}(a). In particular,
the lattice temperature of the Cu atoms forming the nano-filament
tends to be much higher than that of the left and right 
contacts and of the SiO$_2$ matrix. With this respect,
Fig.~\ref{fig:6}(b) distinctly shows that a local hot spot is situated
in the middle and second half of the filament, where the hottest atoms have
temperatures up to 40 K larger than the average of their immediate
surrounding. At high current densities these particles might acquire
enough energy to change site, alter the filament geometry, and destroy
its conductivity in the process. As a side note, it should be
emphasized that classical simulation approaches using the same
relation as above to model $T_{max,avg}$ cannot properly describe the
large, atomic-scale temperature variations of Fig.~\ref{fig:6}(b) and
might underestimate the influence of self-heating.  

\section{Conclusion}\label{sec:conc}
We have studied the electrical and thermal properties of a realistic
conductive bridging random access memory cell from first-principles
with a state-of-the-art solver based on DFT and NEGF. By coupling
electron and phonon transport we could enlighten the mechanisms that
limit the performance of such nano-devices. As possible design
improvement, it appears that bringing closer to each other
the two metallic electrodes that surround the central dielectric
layer, although technologically not evident, would reduce the
nano-filament length and thus prevent electrons from emitting enough
phonons to significantly contribute to self-heating.

\section*{Acknowledgment}
This work was supported by the Werner Siemens Stiftung, by SNF under
Grant No. PP00P2\_159314, by ETH Research Grant ETH-35 15-2, and by a
grant from the Swiss National Supercomputing Centre (CSCS) under
Project s714.

\bibliographystyle{IEEEtran}
%\bibliography{IEEEabrv,references_notitle}
% Generated by IEEEtran.bst, version: 1.13 (2008/09/30)

\newpage

\newpage

\newpage

\newpage

\newpage

\newpage

\end{document}